# Stress - and Magneto-Impedance in $Co_{71-x}Fe_xCr_7Si_8B_{14}$ (x = 0, 2) amorphous ribbons

B. Kaviraj

Magnetism Laboratory, Department of Physics & Meteorology, Indian Institute of Technology, Kharagpur 721302, India. Tel: +91-3222-281629

Email: bhaskar@phy.iitkgp.ernet.in

S.K. Ghatak[*]

Magnetism Laboratory, Department of Physics & Meteorology, Indian Institute of Technology, Kharagpur 721302, India. Tel: +91-3222-283818  Fax: +91-3222-255303

Email: skghatak@phy.iitkgp.ernet.in

---

[*] Corresponding author

**Abstract**

Systematic measurements of stress impedance (SI) and magneto-impedance (MI) have been carried out using Co-rich amorphous ribbons of nominal composition $Co_{71-x}Fe_xCr_7Si_8B_{14}$ (x = 0, 2) at various excitation frequencies and bias fields and at room temperature. The impedance, Z, for both the samples was found to be very sensitive functions of applied tensile stress (up to 100MPa) exhibiting a maximum SI ratio as much as 80% at low frequency ~ 0.1MHz. The nature of variation of impedance, $Z(\sigma)$, changes with the excitation frequency especially at higher frequencies in MHz region where it exhibits a peak. Magnetization measurements were also performed to observe the effects of applied stress and magnetization decreases with the application of stress confirming the negative magnetostriction co-efficient of both the samples. Both the samples exhibited negative magneto-impedance when the variation of Z is observed with the applied bias magnetic field, H. Maximum MI ratio as large as 99% has been observed for both the samples at low fields ~ 27Oe. The impedance as functions of applied magnetic field, Z(H), decreases with the application of stress thus making the MI curves broader. Based on the electromagnetic screening and magnetization dynamics and incorporating the Gilbert and the Bloch-Bloembergen damping and stress dependent anisotropy, the SI has been calculated and is found to describe well the stress and field dependence of impedance of the two samples.

# 1. INTRODUCTION

The magneto-impedance (MI) effect – a large change of impedance induced by d.c. magnetic field is normally observed in a soft ferromagnetic conductor and in most situations the impedance is reduced to a large extent compared to that in a zero field state. The ferromagnetic materials like the amorphous transition metal-metalloid alloys, with the shapes of wires [1-3], ribbons [4-7] and thin films [8-9], are forerunners in exhibiting large negative MI at low frequency [10-13]. The Fe-or Co-based metallic glasses in the form of ribbons or wires are magnetically soft and the relative decrease in impedance in presence of small fields is very large [10-14]. The MI effect has been observed for as-quenched, amorphous as well as in the nanocrystallized materials [15-17]. The MI effect is associated with the field penetration of electromagnetic (e.m) waves within a magnetic metal with high permeability. The impedance of a metal is determined by penetration depth of e.m. field and the penetration depth in turn is a measure of the screening of field. In paramagnetic metals, the screening is solely due to the conduction electrons and an additional component of screening arises from the a.c. magnetization current in ferromagnetic metals. This leads to higher impedance in ferromagnetic state. However, screening of magnetic origin can be altered significantly by changing the magnetic response to a.c field. The response, as measured by the dynamic permeability $\mu$, can be altered by applying an external d.c. magnetic field, external stress, thermal treatment etc. In most cases, a moderate magnetic field increases the penetration depth of e.m. field at r.f range and thereby the impedance is decreased compared to that at zero field [18, 19]. The permeability of a ferromagnetic substance is governed by the magnetic anisotropy energy. The external stress can modify the anisotropy in a material with non-zero magnetostriction coefficient and therefore induce a change in impedance – referred as stress-impedance (SI) [20-22].

In this communication, the SI and MI effects are studied for the sample $Co_{71-x}Fe_xCr_7Si_8B_{14}$ (x= 0, 2) at different excitation frequencies, applied tensile stresses and bias fields. The Co-rich ferromagnetic amorphous alloys have high permeability, low magnetic losses and low magnetostriction constant, and these make the materials suitable for observing both SI and MI effects [23-25]. The model calculation using the Landau-Lifsitz-Gilbert equation has been used to discuss the dependence of impedance of a magnetoelastic film upon various parameters such as excitation frequency, external stress, external magnetic field and orientation of anisotropy and is compared with the experimental investigations.

## 2. EXPERIMENTAL DETAILS

Amorphous ribbons with nominal compositions $Co_{71}Cr_7Si_8B_{14}$ and $Co_{69}Fe_2Cr_7Si_8B_{14}$ were produced using the melt-spinning technique. The sample cross-sections were 6.41x0.0335 and 6.347x0.0325 $mm^2$ respectively. All the samples were cut in 8 cm long pieces for the measurements. All the samples were used in as-quenched state and were aligned with their longitudinal axis perpendicular to the Earth's magnetic field. The samples were placed within a small signal coil of rectangular geometry and with 100 turns. The coil was located at the middle of the sample-length which was larger than the length of the coil in order to reduce the demagnetization effects. The coil was connected to the current terminal of the Impedance Analyzer (Model-HP4294A) where a sinusoidal current amplitude was kept constant at $I_{rms}$ = 20mA. This created an a.c magnetic field of about 196A/m along the axis of the coil which is also along the length of the ribbon. The voltage response of the sample around this field was found to be linear. The frequency of the driving current was scanned from 100KHz to 10MHz. Maximum d.c bias fields up to 27Oe was applied parallel to the direction of exciting a.c field. Magnetization measurements were performed with the help of an a.c magnetometer at 70Hz frequency.

These samples were further subjected to tensile stresses using load and pulley arrangement. Maximum stresses up to 100MPa were applied longitudinally depending upon the cross-section of the ribbon. The real and imaginary components of impedance Z = R+jX where X= ωL (L being the inductance of the sample) were measured across the signal coil with the help of Impedance Analyzer. The resistance and reactance of the empty coil and test leads were subtracted and only the change in impedance due to the sample was taken into account. All measurements were performed at room temperature. The stress-impedance ratio and the magneto-impedance ratio has been expressed as

$$\delta Z_\sigma = \frac{Z(\sigma) - Z(0)}{Z(0)} \times 100 \text{ and } \delta Z_H = \frac{Z(H) - Z(0)}{Z(0)} \times 100.$$

## 3. RESULTS

### 3.1. Sample $Co_{71}Cr_7Si_8B_{14}$ (x = 0)

Fig. 1 shows the excitation frequency response of the resistive (R) and reactive (X) components of impedance for the sample $Co_{71}Cr_7Si_8B_{14}$ as functions of different external tensile stresses. At all stresses, R and X are low at low frequencies but increase monotonically at higher frequencies. The reactive part is greater than the resistive part. The applied tensile stress (σ) decreases the magnitude of R progressively as the stress is increased from 0MPa to 100MPa. The frequency response of the reactive component of Z on the other hand exhibits a 'cross-over' in high frequency region (> 1MHz) at σ = 20MPa where it crosses the 'zero-stress' curve indicating that the values of X at 20MPa stress are greater than those at 'zero-stress' in this frequency region. But at σ = 100MPa, the values of X are lower than those at σ = 0MPa. These will be clear from the results of stress impedance measurements that are presented subsequently.

In Fig. 2, we show the stress impedance results for the sample $Co_{71}Cr_7Si_8B_{14}$ measured at different excitation frequencies of 0.1MHz, 1MHz and 10MHz. Here the percentage change in impedance $\delta Z_\sigma$ has been plotted as functions of different tensile stresses. At low frequencies, the impedance is maximum at zero stress and falls off sharply with the increase of stress. At a higher frequency (10MHz), the impedance increases with stress and exhibits a maximum at σ ~ 20MPa but at higher stresses $\delta Z_\sigma$ becomes negative. The maximum relative change in SI decreases from 80% (0.1MHz) to around 40% (10MHz) with the increase in frequency in the range of applied stresses.

The impedance (Z) has also been measured at different d.c biasing fields (H) at 1MHz frequency and the results for $\delta Z_H = \frac{(Z(H) - Z(0))}{Z(0)} \times 100$ versus H are depicted in Fig. 3 for σ = 0, 32, 65, 90MPa. The impedance of the sample exhibits a maximum at zero d.c field and decreases monotonically with H, thus exhibiting a negative magneto-impedance phenomenon. In the absence of stress (σ = 0), the impedance of the sample has been found to drop from 36Ω (H = 0) to 0.02Ω (H = 27Oe) with the application of field (see inset of Fig. 3). The impedance saturates to very low values for H > 20Oe. The maximum change in magneto-impedance (MI) ratio with respect to zero-field was found to be nearly 99% for all values of applied stresses. With the increase of applied stress, the MI response becomes flatter and it saturates at higher values of applied d.c field. At higher frequencies, the observed field dependence of Z was found to remain the same but the maximum MI ratio decreased at higher values of stresses.

### 3.2. Sample $Co_{69}Fe_2Cr_7Si_8B_{14}$ (x = 2)

The frequency response of the resistive and reactive components of impedance for this particular ribbon is shown in Fig. 4. Fig. 4 shows the monotonic increase of resistive and reactive components of impedance at high frequencies. The 'cross-overs' with the $\sigma = 0$ curve suggests that within this stress interval, the relative change in SI is positive.

Fig. 5 depicts the relative change in SI for the sample $Co_{69}Fe_2Cr_7Si_8B_{14}$ as functions of different stresses at different excitation frequencies (0.1, 1 and 10MHz). Comparing with the SI response of the previous sample in Fig. 2, we notice that in this sample the maximum relative changes in SI are lower at all the chosen range of frequencies. The maximum change is around 70% and 47% at 0.1 and 1MHz respectively. At 1MHz and 10MHz, the SI% becomes positive (hence exhibits a peak) at finite intervals of stresses and thereafter decreases. The stress value at which SI% shows a maxima increases from $\sigma \sim 9MPa$ at 1MHz to $\sigma \sim 30MPa$ at 10MHz. Also compared to the previous sample, we note that in this sample, the SI% remains positive for a larger extent of stress at high frequencies $\sim 10MHz$. It is positive for the entire range of stress.

In Fig. 6, the field dependence of magneto-impedance has been plotted as functions of different external stresses and at an excitation frequency of 1MHz. This sample also exhibits negative magneto-impedance for all values of applied stresses. Maximum MI ratio up to 99% has been observed for all values of stresses. We note that the zero-field impedance values for σ = 32MPa is larger than the corresponding value for σ = 0MPa, thus the sample exhibiting positive SI ratio within this interval of applied stress (see inset). The zero-field impedance values for higher stresses (σ >32MPa) are lower than that of zero-stress. This also supports the observations in Fig. 5 where the SI ratio for 1MHz frequency remained positive up to σ ~30MPa and becomes negative thereafter with the increase of stress.

The magnetization measurements of both the samples have been performed by fluxmetric method and at a frequency of 70Hz. Different external stresses also have been applied to study the effects of the change in the magnetization curves. They are depicted in the figures below (Fig. 7 and Fig. 8). Figs. 7 and 8 show that magnetization for both the samples decreases with the application of external stress.

## 4. Discussions

From Fig. 2 and and Fig. 5 we note that the impedance decreases with the application of stress. This decrease is monotonic at frequencies up to 1MHz for the sample with $x = 0$ and even lower up to KHz region for the sample with $x = 2$. This is because the Co-rich samples possess very small (near to zero) and negative magnetostriction coefficient. Increasing stress (tensile) in them promotes the growth of domains in the perpendicular direction, which in our geometry is the direction transverse to the long axis of the ribbon (and also transverse to the exciting a.c field). This causes a reduction in the longitudinal permeability of the ribbon and hence a reduction of impedance. Moreover at higher excitation frequencies, the sharpness of SI curves decreases. This is attributed to the decrease of permeability at high frequencies.

The magnetization for both the samples decreases with the application of external stress (see Figs. 7 and 8) because of the negative magnetostriction coefficient of the samples. The calculated values of saturation magnetization ($M_s$) were $1.77 \times 10^5$ A/m and $2.47 \times 10^5$ A/m for x = 0 and x = 2 ribbons respectively. Such high values of $M_s$ are close to those that are reported for Co-rich amorphous alloys [26-27]. Comparing the two samples, we notice that the magnitudes of magnetization are higher for the sample $Co_{69}Fe_2Cr_7Si_8B_{14}$ (Fig. 8) than the one having composition $Co_{71}Cr_7Si_8B_{14}$ (Fig. 7). This is because of the fact that the magnetostriction coefficient ($\lambda_s$) of the pure Cobalt sample (x=0) is higher than that of $Co_{69}Fe_2Cr_7Si_8B_{14}$ (x = 2). The estimated $\lambda_s$ from the slope of anisotropy energy ($E_k$) versus applied stress ($\sigma$) curves was found to be $-0.63 \times 10^{-6}$ and $-0.24 \times 10^{-6}$ for x = 0 and 2 respectively. These values were close to the measured values of $\lambda_s$ for Co-rich alloys [28, 29]. (Subtle variations in the measured values of $\lambda_s$ may arise from the internal stresses that are produced during the quenching process of the amorphous material). Since the anisotropy of amorphous magnetic materials is predominantly magnetostrictive in nature, the pure Cobalt sample possesses higher anisotropy energy and hence lower values of susceptibility than that of $Co_{69}Fe_2Cr_7Si_8B_{14}$.

The comparison of the SI response of the two samples is shown in Fig. 9. The results show that the SI response of the sample with x = 2 is broader compared to the sample with x = 0 in both the frequency regimes. The difference in broadness of the SI curves increases as we go to the higher frequency. The results can be explained by the fact that the magnetostriction coefficients of both the samples are negative but lower in magnitude and close to zero in the sample with x = 2. The anisotropy energy in amorphous ribbons arises mainly from the coupling of the internal stresses with the magnetization due to the magnetoelastic effect and is lower in the sample x = 2 compared to that in x = 0. A transverse component of magnetic anisotropy inadvertently exists in the as-cast sample [9]. Since in a negative magnetostrictive sample, the application of stress promotes the growth of domains perpendicular to the direction of stress, the rate of decrease of permeability (longitudinal) must be greater for the sample with higher negative magnetostriction. Hence the fall of impedance with stress is also sharper for the sample with higher negative magnetostriction giving rise to sharper SI curves.

## 5. MODEL AND FORMULATION

It is well known that the AC current is not homogenous over the cross-section of the magnetic conductor due to the screening of e.m field. The screening is governed by the Maxwell equations along with the magnetization dynamics. In soft ferromagnetic materials *M* becomes a non-linear function of *H* and this leads to nonlinear coupled equations for *M* and *H*. Assuming the linear magnetic response of the material, the above coupled equation can be solved and inhomogeneous distribution of the current is then characterized by the skin depth

$$\delta = \sqrt{\frac{1}{\sigma_c f \mu_{eff}}} \qquad (1)$$

where 'f' is the frequency of the AC current, $\sigma_c$ is the electrical conductivity and $\mu_{eff}$ is the effective permeability of the magnetic film. In magnetic films, the permeability $\mu_{eff}$ depends upon the frequency f, the amplitude of the bias magnetic field, the applied stress, etc. In the experimental condition, the electric and magnetic fields of e.m. excitation are perpendicular and parallel to the ribbon length respectively. This is in contrast to the usual measuring situation where the excitation current flows along the length of the ribbon and hence the field directions are interchanged. Assuming the magnetic field along z-direction and electric field $\vec{e}_y$ along y-direction the induced voltage, $V_s$, across the signal coil around the sample is given by [18] $V_s = \int \vec{e}_y . \vec{dl} = ZI_0$ where $I_0$ is the amplitude of a.c current in the coil. The impedance $Z$ of a long magnetic ribbon of thickness '2d' is given by:

$$Z = -i\omega L_0 \mu \qquad (2)$$

where $\mu = \mu_{eff} \dfrac{tanh(kd)}{kd} \qquad (3)$

with k = (1+i) /δ  (4)

and $L_0 = \dfrac{n^2 A_c}{2l} \mu_0$ is the inductance of the empty coil and $\mu_{eff}$ is the effective permeability.

The permeability $\mu_{eff}$ of the ferromagnetic material is a complex quantity due to magnetic relaxation and alters the impedance of the sample in a non-linear way in presence of external stress or magnetic field. As the permeability is a measure of magnetic response, it is necessary to consider the magnetization dynamics in presence of small excitation field and external parameters and then to estimate the permeability and its variation with the parameters. The magnetization dynamics of ferromagnetic material in macroscopic scale is customarily described by the Landau-Lifsitz-Gilbert equation:

$$\dot{\vec{M}} = \gamma\left(\vec{M} \times \vec{H}_{eff}\right) - \frac{\alpha}{M_s}\left(\vec{M} \times \dot{\vec{M}}\right) - \frac{1}{\tau}\left(\vec{M} - \vec{M}_0\right) \qquad (5)$$

Here $M$ is the magnetization, $\gamma$ is the gyromagnetic ratio, $M_s$ the saturation magnetization, $H_{eff}$ is the effective field and $M_0$ the equilibrium magnetization. The first term in right hand side of equation (5) is torque acting on $M$ due to $H_{eff}$, the second term is the Gilbert damping term with damping coefficient $\alpha$. The last term is referred as the Bloch Bloembergen damping with relaxation time $\tau$. This does not preserve the magnitude of macroscopic magnetization, as is required for an ideal ferromagnet, and is used to describe the relaxation processes in materials with imperfect ferromagnetic order (such as amorphous and nanocrystalline alloys or crystals with some structural defects). It has been proved that the choice of the particular damping term substantially influences the imaginary part of effective permeability and consequently the magnitude of magneto-impedance effect [30]. The effective field $H_{eff}$ can be written as:

$$\vec{H}_{eff} = \vec{H} + \vec{H}_a + \vec{H}_\sigma \qquad (6)$$

where the exchange coupling field and the demagnetizing field have been neglected to simplify the computation. The corresponding fields are defined as follows: **H** is the sum of applied d.c bias field and exciting a.c field. The uniaxial anisotropy field is

$$H_a = \frac{2K_u}{\mu_0 M_s^2} \vec{e}_a\left(\vec{e}_a \bullet \vec{M}\right) = \frac{H_k}{M_s} \vec{e}_a\left(\vec{e}_a \bullet \vec{M}\right) \tag{7}$$

where $e_a$ is the unit vector along the easy axis. $K_u$ is the uniaxial anisotropy constant and

$$H_k = \frac{2K_u}{\mu_0 M_s}$$

The applied stress effective field is

$$\vec{H}_\sigma = \frac{H_{\sigma 1}}{M_s} \vec{e}_\sigma\left(\vec{e}_\sigma \cdot \vec{M}\right) \tag{8}$$

where $e_\sigma$ is the unit vector along the applied stress direction and

$$H_{\sigma 1} = \frac{3\lambda\sigma}{\mu_0 M_s} \tag{9}$$

where $\lambda$ is the magnetostriction coefficient. In presence of low amplitude driving current *I*, the excitation a.c field *h* is much smaller than the other magnetic fields. Therefore, the induced magnetization **m** is small and the deviation of **M** from its equilibrium orientation $M_0$ is also small. So one can assume $H_{eff} = H_{eff0} + h_{eff}$ and $M = M_0 + m$ and the a.c component of the vectors varies as

**h, $h_{eff}$, m** $\alpha$ $e^{i\omega t}$ (10)

where $\omega = 2\pi f$ is the circular frequency of the a.c current.

From Eqs. (6), (7), (8) and (10) we get

$$H_{eff0} = \vec{H} + \frac{H_k}{M_s}\vec{e}_a\left(\vec{e}_a \bullet \vec{M}_0\right) + \frac{H_{\sigma 1}}{M_s}\vec{e}_\sigma\left(\vec{e}_\sigma \bullet \vec{M}_0\right) \tag{11}$$

$$\vec{h}_{eff} = \vec{h} + \frac{H_k}{M_s}\vec{e}_a\left(\vec{e}_a \cdot \vec{m}\right) + \frac{H_{\sigma 1}}{M_s}\vec{e}_\sigma\left(\vec{e}_\sigma \cdot \vec{m}\right)$$

$$= \vec{h} + \vec{h}_a \qquad (12)$$

Substituting Eqs.(10)-(12) into (5) and then rewriting (5)

$$\frac{i\omega^*}{\gamma}\vec{m} + \left(\frac{i\alpha\omega}{\gamma}\frac{\vec{M}_0}{\vec{M}_s} + \vec{H}_{eff0}\right) \times \vec{m} = \left(\vec{M}_0 \times \vec{h}_{eff}\right) \qquad (13)$$

Here $\omega^* = \omega - i/\tau$, where $\tau$ is related to the relaxation frequency $\omega_0$, by $\omega_0 = 1/\tau$. The a.c components of magnetization can be computed from equation (13) and the effective permeability can be obtained by

$$\mu_{eff} = \left(\frac{m_z}{h_z} + 1\right) \qquad (14)$$

The impedance of the magnetoelastic film can then be obtained from the Eqs. (1), (2) and (14). For solution of equation (13), it is necessary to determine the components $M_\theta$. In the absence of a.c magnetic field, the equilibrium magnetization, as follows from equation (5), is determined by

$$\left(\vec{M}_0 \times \vec{H}_{eff}\right) = 0 \qquad (15)$$

With this equation, the equilibrium angle θ could be obtained when $H$, $H_a$ and $H_\sigma$ are given.

The magnetic ribbon is modeled as a system with a uniform uniaxial in-plane magnetic anisotropy oriented at an angle $\theta_0$ with respect to the z-axis (Fig.10). The stress σ or d.c field ($H$) is assumed to be applied along z-axis and the magnetization $M_0$, in absence of a.c. field, lies in the film plane at an angle θ with respect to the z-axis. The equilibrium angle θ is first obtained for different stresses from equation (15) and these values are used to calculate $\mu_{eff}$ (equation 14) and hence the impedance Z as functions of frequency, stress and magnetic field is determined. Considering the empirical relation between bulk magnetostriction and magnetization [31] we write

$$\lambda = \gamma_1 M^2 \qquad (16)$$

where $\gamma_1$ depends upon the stress. The stress dependence of the magnetostriction curve $\lambda(M,\sigma)$ can be described in terms of $\gamma_1$ and can be expressed in Taylor series in powers of σ [31] as:

$$\gamma_1(\sigma) = \gamma_1(0) + \sum_{n=1}^{\infty} \frac{\sigma^n}{n!} \gamma_1^n(0) \qquad (17)$$

where $\gamma_1^n(0)$ is the nth derivative of $\gamma_1$ with respect to stress at σ = 0.

## 6. NUMERICAL RESULTS AND DISCUSSION

Based on the above theoretical model and using typical values of parameters $\theta_0 = 50^0$, $H_a = 200 A/m$, $\alpha = 20$, $M_s = 6.5 \times 10^5$ A/m and $\omega_0 = 10^6$ rad/s, we first present the frequency response of real and imaginary components of impedance as functions of different applied stresses. This is shown in Fig. 11. The values of $\gamma_1(0) = -7 \times 10^{-18} A^{-2} m^2$ and $\gamma_1'(0) = 1 \times 10^{-29} A^{-2} m^2 Pa^{-1}$ were used in all the cases. These values are typical for soft ferromagnetic materials [31]. The value of $H_a$ chosen was close to the anisotropy fields of FeCoSiB alloys [26].

From Fig. 11, we note that the values of Z for $\sigma = 10MPa$ are higher than those for $\sigma = 0MPa$ at the whole interval of frequency. For higher values of stresses ($\sigma > 10MPa$), Z decreases monotonically with $\sigma$ at low frequencies. At higher frequencies, the impedance curves corresponding to a particular value of stress crosses the $\sigma = 0MPa$ curve thereby making $Z(\sigma) > Z(0)$ implying that the SI ratio is positive within these intervals of frequency. Such 'cross-overs' with the $\sigma = 0$ curve in the frequency dependence of Z has also been observed experimentally in both the samples of $Co_{71-x}Fe_xCr_7Si_8B_{14}$ (x = 0, 2) (Fig. 1 and Fig. 4). The inset shows the complex behavior of susceptibility as functions of different frequencies and applied stresses. The real component of susceptibility ($\chi'$) initially remains constant with frequency but decreases to a large extent at higher frequencies. Further the values of $\chi'$ corresponding to 10MPa are higher than those of 0MPa for all frequencies which may be correlated with the behavior of Z. The imaginary component of susceptibility ($\chi''$) increases from very low values at low frequencies, exhibits maxima at intermediate frequencies and then decreases monotonically. With the increase of stress, the peak value of imaginary susceptibility ($\chi''_{max}$) also becomes a function of applied stress and it gets shifted to higher frequencies with the application of stress.

We now discuss the stress-impedance effect as a function of different orientations of anisotropy field $H_a$. From equation (18) in our model which governs the equilibrium position of **M**, we see that $\theta_0$ determines the orientation of the magnetization vector. Fig. 12 shows the variation of relative change in impedance as functions of applied stresses and for different values of $\theta_0$ at 100KHz frequency, where $\theta_0$ is the angle between the anisotropy field $\boldsymbol{H_a}$ and the direction of stress (z-axis).

From Fig. 12, we find that the orientation of anisotropy has a major role to play in SI effect. The impedance response for low values of $\theta_0$ (up to $50^0$) is entirely different from those of still higher values. Up to $\theta_0 = 50^0$, Z increases at lower values of stress, exhibits a maxima and then decreases monotonically at higher stresses. The position of impedance maxima shifts to lower values of stress as $\theta_0$ is increased to $50^0$. This behavior changes entirely when $\theta_0 > 50^0$ where Z decreases monotonically with the applied stress. The sharpness of fall of Z also increases with $\theta_0$ in the regime of $\theta_0 > 50^0$.

In Fig. 13, we depict the variation of relative change in magneto-impedance with external biasing magnetic field, H (normalized with respect to the anisotropy field $H_a$ = 200A/m) at different applied stresses and at 1MHz frequency. The parameter values used to simulate the field dependence of impedance are $\theta_0 = 50^0$, $H_a = 200$A/m, $\alpha = 20$, $M_s$ = $6.5 \times 10^5$A/m, $\omega_0 = 10^6$ rad/s, $\gamma_1(0) = -7 x 10^{-18} A^{-2} m^2$ and $\gamma_1'(0) = 1 x 10^{-29} A^{-2} m^2 Pa^{-1}$. The impedance exhibits a maximum at zero bias field and decreases with the increase of bias field, thus exhibiting negative magneto-impedance. The results show a maximum MI ratio of about 99% for all stresses. With the increase of σ, the MI response becomes flatter and it saturates at higher fields. Such dependences of Z on applied magnetic field are very similar to those observed experimentally (Figs. 3 and 6). The inset shows the variation of relative change in impedance with the scaled field $H/H_{1/2}$, $H_{1/2}$ being the field where ($\Delta Z/Z$) reduces to half its maximum, at different stresses obtained from the theoretical model and has been compared with the experimental data points of Fig. 6 for the sample $Co_{69}Fe_2Cr_7Si_8B_{14}$ (x=2). It shows that except for the 0MPa theoretical line, all the theoretical and experimental curves almost collapse into a single curve implying a good agreement between the theory and experiment.

The stress-impedance behavior, in particular the dependence of Z on applied stresses at various excitation frequencies obtained from the theoretical model is shown in Fig. 14. Fig. 14 depicts the frequency dependence of SI effect at two different values of $\theta_0$. Fig. 14(a) correspond to $\theta_0 = 60^0$ and 14(b) to $\theta_0 = 50^0$. All the other parameters have been kept fixed. The results clearly depict the formation of peaks in the SI curves with the increase in frequency. This has also been borne out by the experiments (Fig. 2 and Fig. 5). From Fig. 14(b), we note that the SI response for 10MHz frequency is positive for the entire interval of stress. This has a resemblance with the response of the sample $Co_{69}Fe_2Cr_7Si_8B_{14}$ (x=2) (Fig. 5). On the other hand, according to the results of Fig. 14 (a), the SI ratio exhibits a monotonic decrease with σ at low frequencies but exhibits a peak at a higher frequency of 10MHz. The stress corresponding to this maximum is close to 20MPa which also resembles the SI behavior for the sample $Co_{71}Cr_7Si_8B_{14}$ (x=0) at 10MHz frequency where the peak of SI ratio was obtained at 20MPa (Fig. 2). So Fig. 14(a) and (b) resembles the observed SI behavior of $Co_{71}Cr_7Si_8B_{14}$ (x=0) and $Co_{69}Fe_2Cr_7Si_8B_{14}$ (x=2) respectively. The results (Fig. 14) show a good agreement between the theoretical model and the observed experimental results.

## 7. CONCLUSIONS

Systematic measurements of stress-impedance and magneto-impedance for the samples $Co_{71-x}Fe_xCr_7Si_8B_{14}$ (x = 0, 2) lead us to conclude

(i) the variation of impedance with stress is a function of excitation frequency. At low frequency ~ 100KHz, impedance decreases monotonically with the applied stress but at high frequencies at or above 1MHz, the impedance rises to a maximum and then decreases at higher stress.

(ii) due to the presence of very small and negative magnetostriction coefficient in both the alloys, very large and sharp changes in magneto-impedance have been observed with the application of very low biasing d.c fields (up to 27Oe). Maximum changes in magneto-impedance as large as 99% have been observed in both the alloys and the sharpness of magneto-impedance response decreases with the increase of applied stress.

(iii) a theoretical model for the formulation of impedance as functions of different parameters (excitation frequency, external magnetic field, angle of anisotropy, external stress, etc) has been constructed based on the solution of Landau-Lifsitz-Gilbert equation of motion and is found to agree very well with the experimental results. The maximum observed SI ratios especially at lower excitation frequencies for both the samples are higher than those predicted by the model. This is because the model was based on a single-domain and hence we neglected the susceptibility due to domain wall displacements. Only the response due to magnetization rotation was considered.

## ACKNOWLEDGEMENTS

The authors are grateful to Dr. A. Mitra for providing the samples and S. Ghosh for technical help.

**Figure Captions**

Fig 1. Frequency response of resistive (a) and reactive (b) components of impedance as functions of different tensile stresses for the sample $Co_{71}Cr_7Si_8B_{14}$ (x =0).

Fig. 2. The variation of relative change in impedance with stress denoted by $\delta Z_\sigma = \frac{Z(\sigma) - Z(0)}{Z(0)} \times 100$ for the sample $Co_{71}Cr_7Si_8B_{14}$ (x = 0) at 0.1, 1 and 10MHz frequency.

Fig.3. The relative change in magneto-impedance (%) denoted by $\delta Z_H = \frac{(Z(H) - Z(0))}{Z(0)} \times 100$ as functions of different external stresses for the sample $Co_{71}Cr_7Si_8B_{14}$ (x = 0) at 1MHz frequency. Inset shows the field dependence of absolute values of impedance.

Fig 4. Frequency response of resistive (a) and reactive (b) components of impedance as functions of different tensile stresses for the sample $Co_{69}Fe_2Cr_7Si_8B_{14}$ (x =2).

Fig.5. The variation of relative change in impedance $\delta Z_\sigma = \frac{Z(\sigma) - Z(0)}{Z(0)} \times 100$ with stress for the sample $Co_{69}Fe_2Cr_7Si_8B_{14}$ (x = 2) at 0.1, 1 and 10MHz frequency.

Fig.6. The relative change in magneto-impedance (%) denoted by $\delta Z_H = \frac{(Z(H) - Z(0))}{Z(0)} \times 100$ as functions of different external stresses for the sample $Co_{69}Fe_2Cr_7Si_8B_{14}$ (x = 2) at 1MHz frequency. Inset shows the field dependence of absolute values of impedance.

Fig. 7. Magnetization Curves for the sample $Co_{71}Cr_7Si_8B_{14}$ (x = 0) as functions of different tensile stresses.

Fig. 8. Magnetization Curves for the sample $Co_{69}Fe_2Cr_7Si_8B_{14}$ (x = 2) as functions of different tensile stresses.

Fig. 9. Comparison of SI response of the samples(x = 0, 2) at excitation frequencies of 100KHz (a) and 10MHz (b).

Fig. 10. Schematic view of magnetoelastic film showing different directions of anisotropy fields and stress.

Fig11. Frequency dependence of Z as functions of applied stresses. The values of parameters taken are $\theta_0 = 50^0$, $H_a = 200$A/m, $\alpha=20$, $M_s = 6.5 \times 10^5$ A/m, and $\omega_0 = 10^6$ rad/s. Inset shows the frequency response of susceptibility at different applied stresses with the same parameters.

Fig. 12. Dependence of SI effect on direction of uniaxial anisotropy ($\theta_0$) at 100KHz frequency. The values of parameters taken are $\theta_0 = 50^0$, $H_a = 200$A/m, $\alpha=20$, $M_s = 6.5 \times 10^5$ A/m and $\omega_0 = 10^6$ rad/s.

Fig. 13. Variation of relative change in magneto-impedance [$(\Delta Z/Z)\% = 100 \times (Z(H)-Z(0)/Z(0))$] with normalized field $H/H_a$ ($H_a = 200$A/m) at different applied stresses and at 1MHz frequency. The values of various parameters are $\theta_0 = 50^0$, $H_a = 200$A/m, $\alpha = 20$, $M_s = 6.5 \times 10^5$A/m, $\omega_0 = 10^6$ rad/s $\gamma_1(0) = -7 \times 10^{-18} A^{-2} m^2$ and $\gamma_1'(0) = 1 \times 10^{-29} A^{-2} m^2 Pa^{-1}$. The inset shows the impedance variation with the scaled field $H/H_{1/2}$ ($H_{1/2}$ being the field where ($\Delta Z/Z$) reduces to half its maximum). The solid lines represent the theory whereas the data points correspond to the experiment for the sample $Co_{69}Fe_2Cr_7Si_8B_{14}$ (x=2).

Fig. 14. The stress dependence of relative change in impedance ($\Delta Z/Z$) % as functions of excitation frequencies obtained from the theoretical model with $\theta_0 = 60^0$ (a) and $\theta_0=50^0$ (b). All the other parameters $H_a = 200$A/m, $\alpha = 20$, $M_s = 6.5 \times 10^5$A/m, $\omega_0 = 10^6$ rad/s, $\gamma_1(0) = -7 \times 10^{-18} A^{-2} m^2$ and $\gamma_1'(0) = 1 \times 10^{-29} A^{-2} m^2 Pa^{-1}$ were kept fixed.

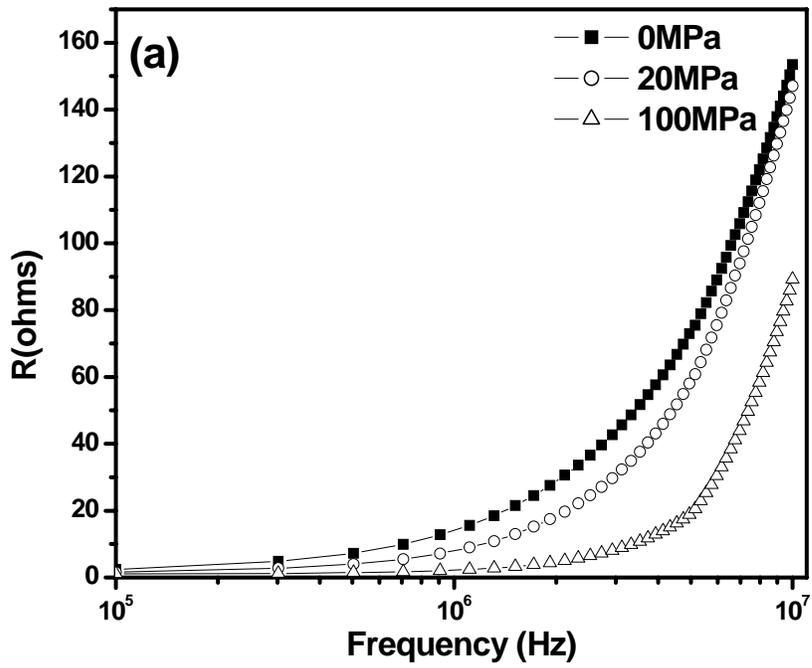
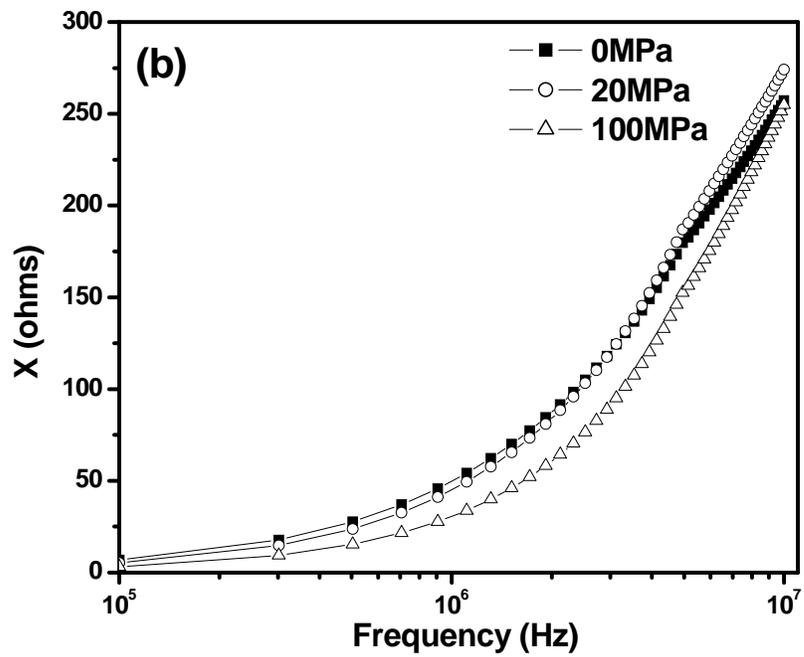

**Fig. 1. B. Kaviraj et al.**

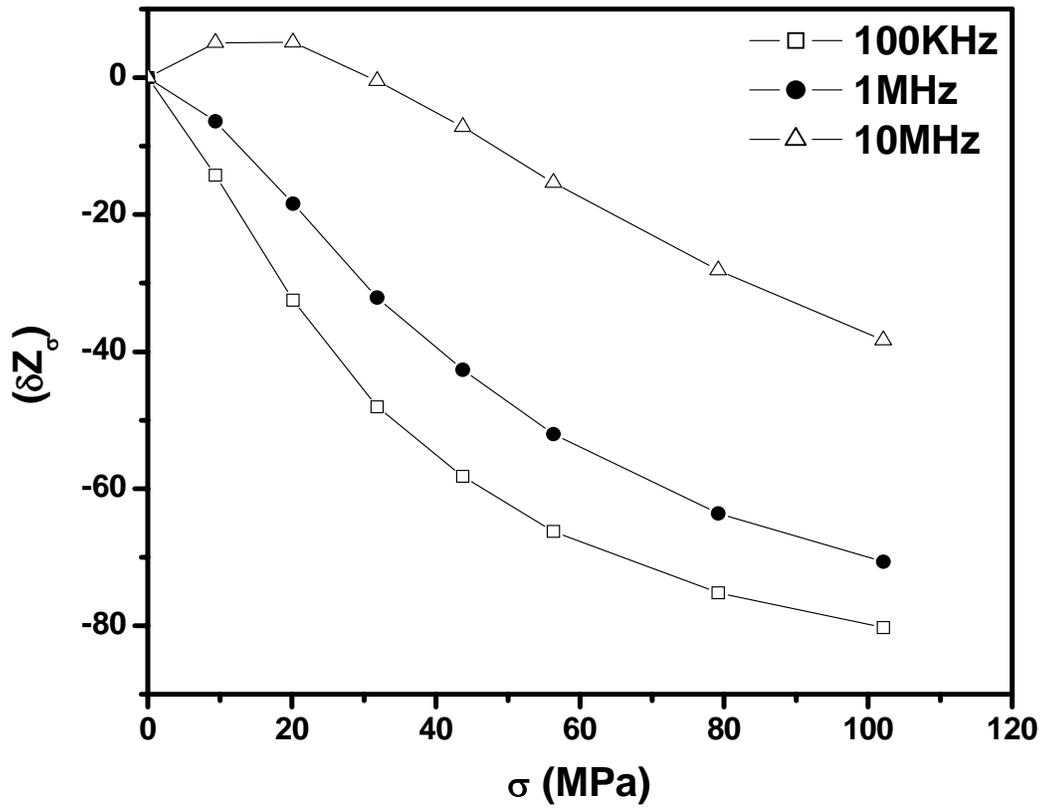

Fig. 2. B. Kaviraj et al.

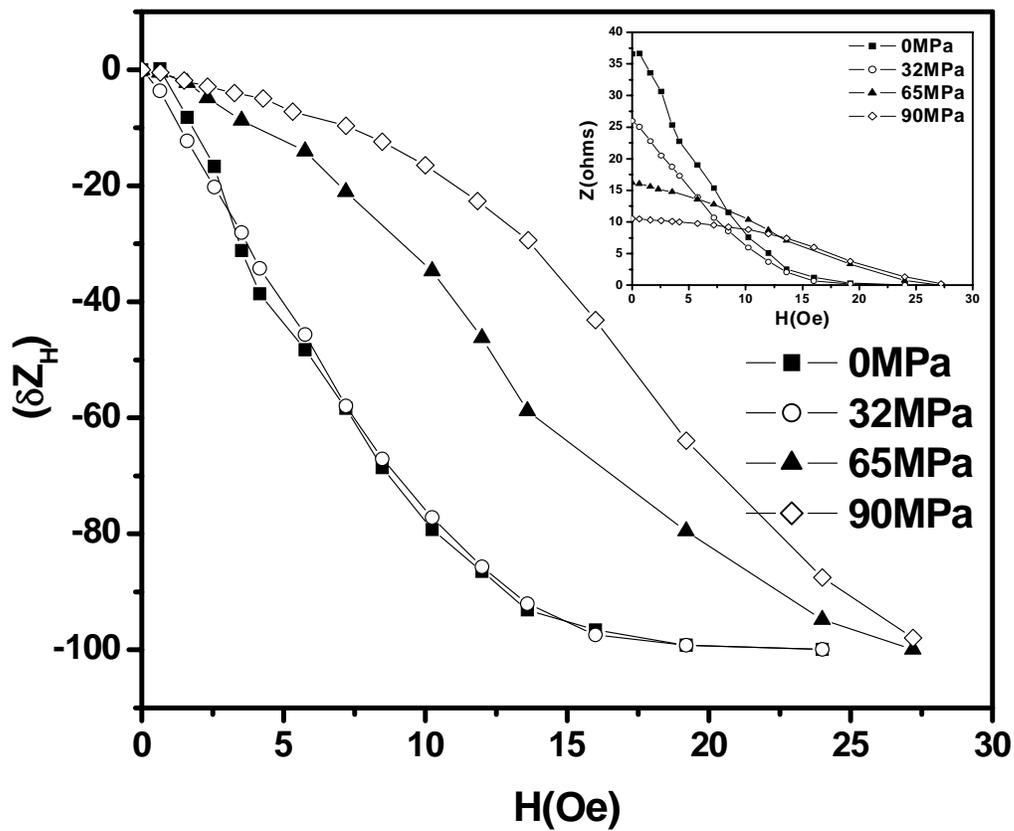

Fig. 3. B. Kaviraj et al.

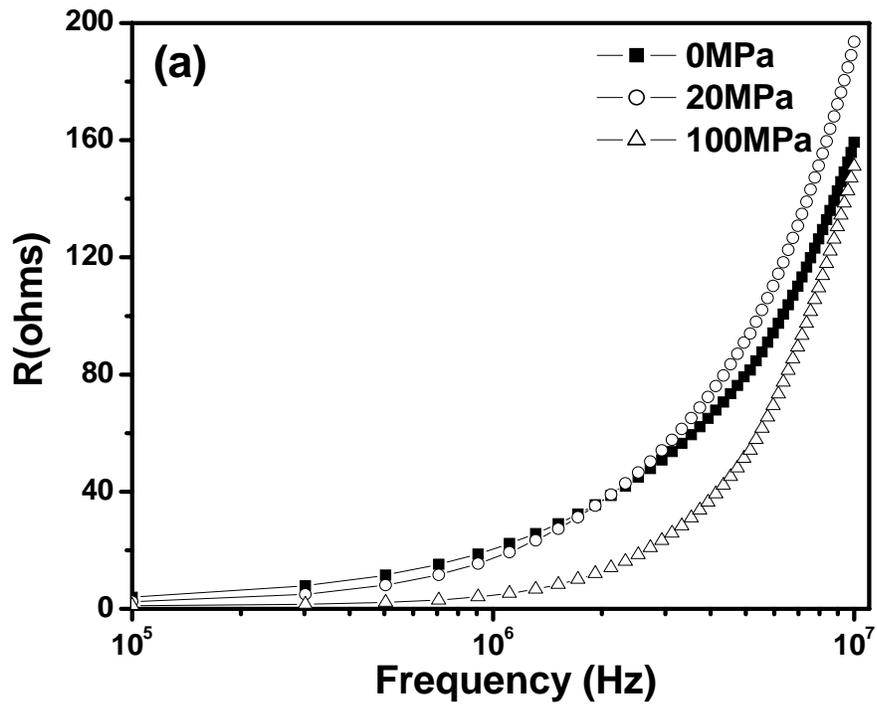

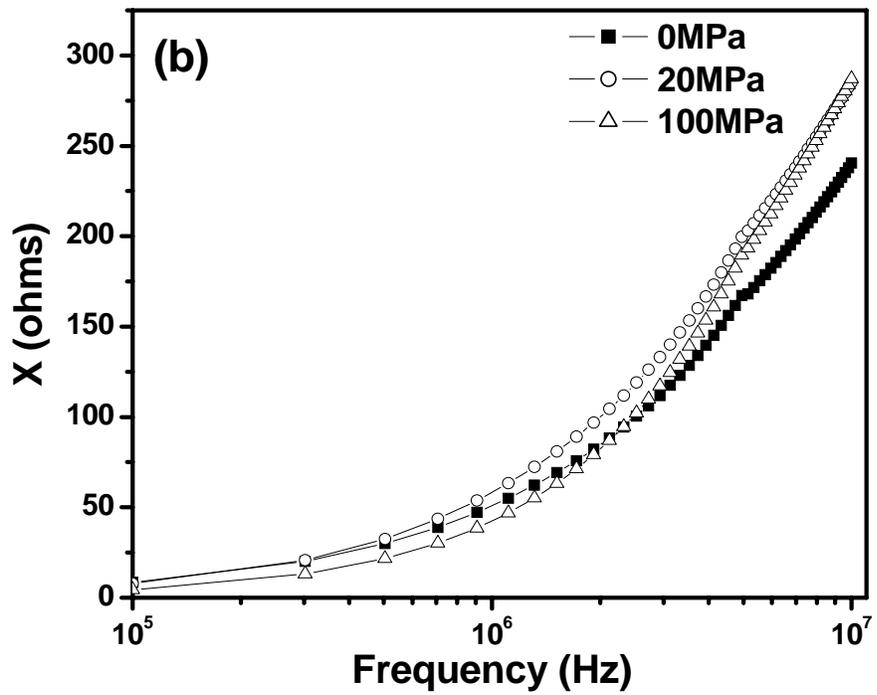

**Fig. 4. B. Kaviraj et al.**

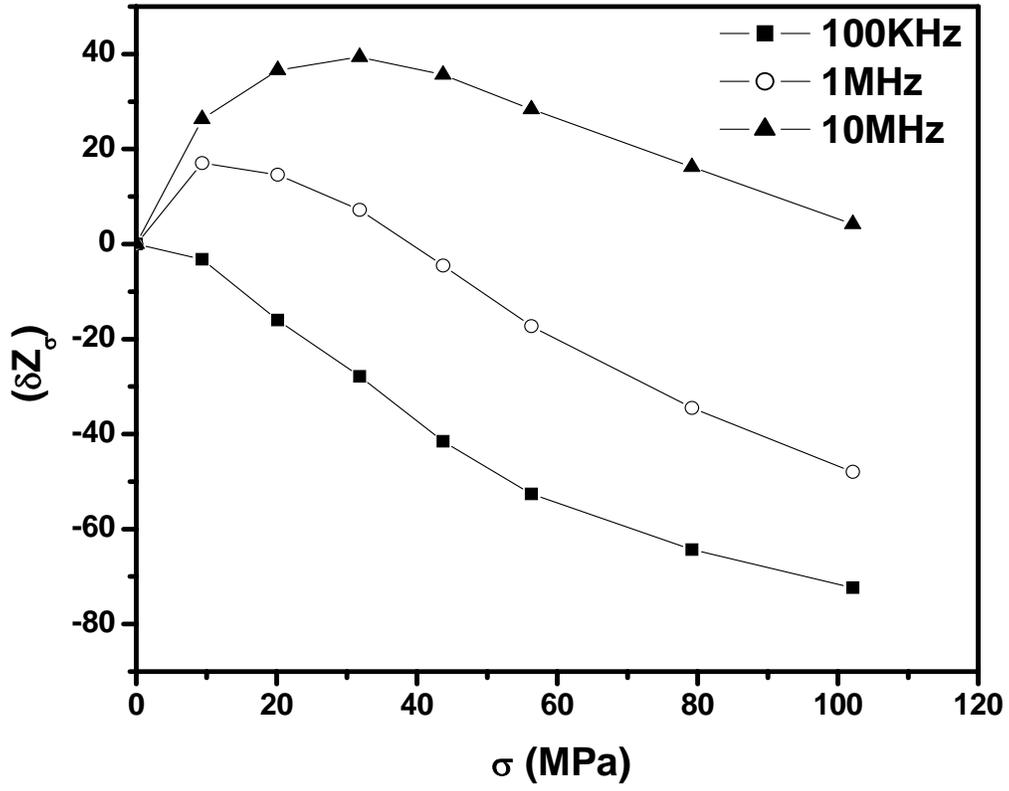

Fig. 5.  B. Kaviraj et al

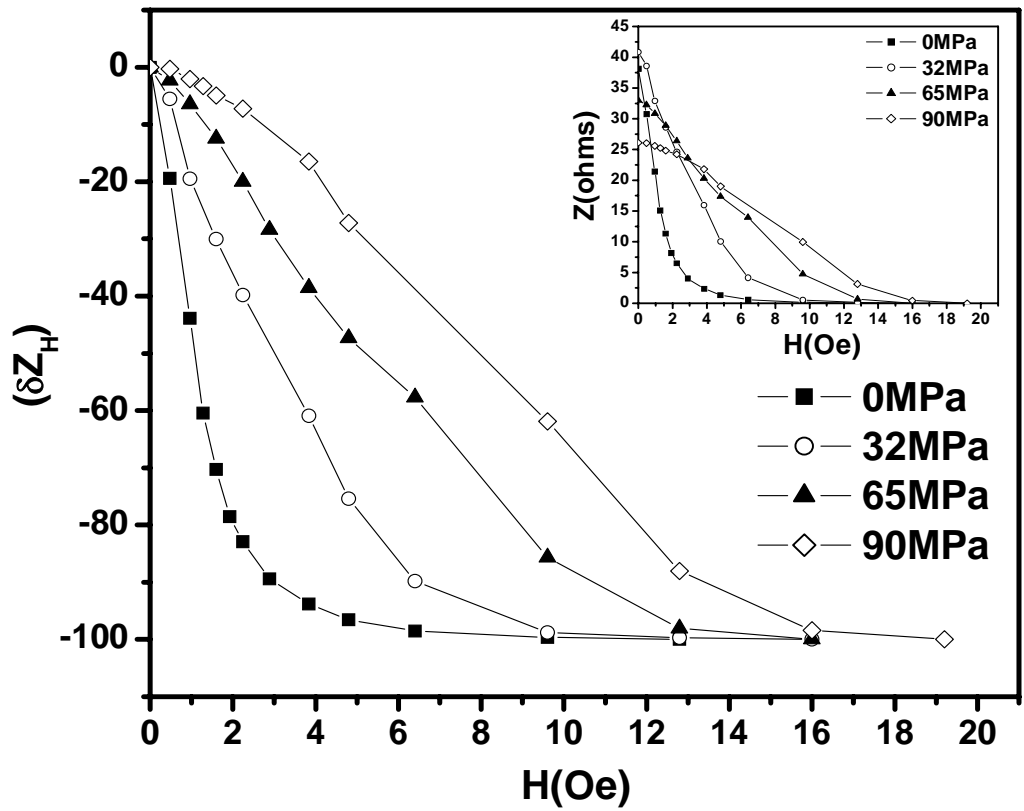

Fig. 6. B. Kaviraj et al.

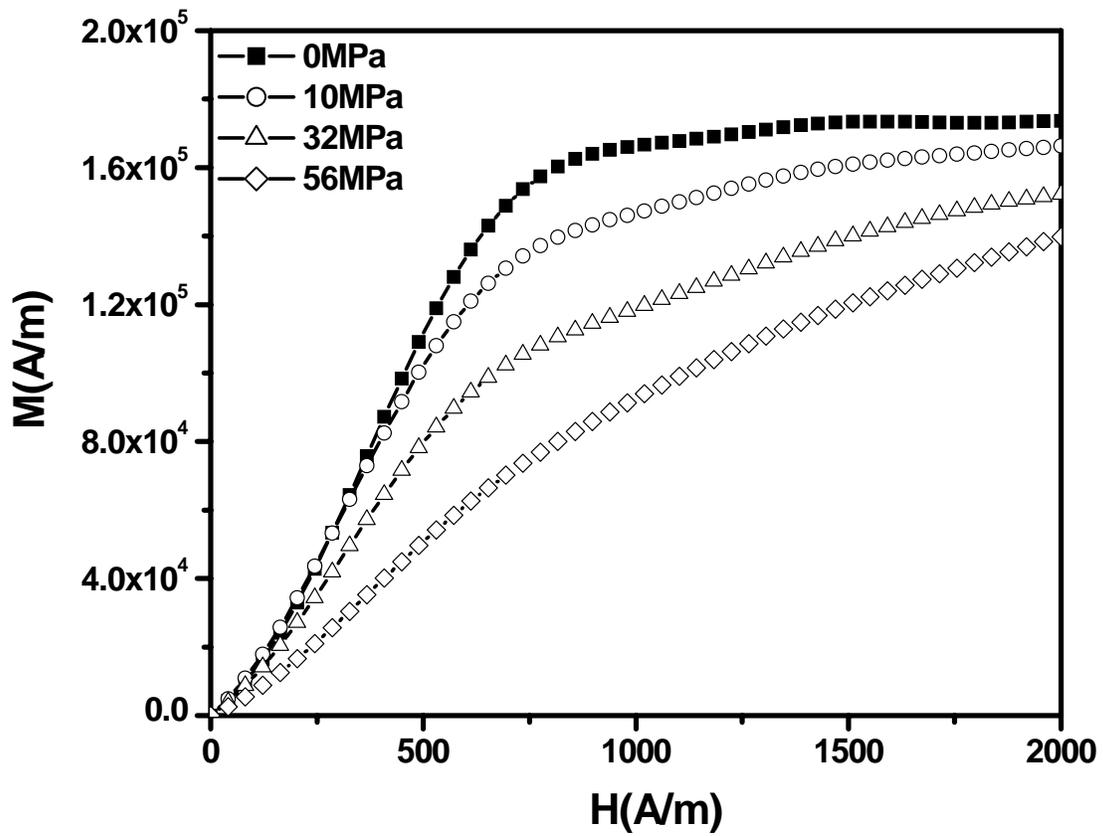

Fig. 7. B. Kaviraj et al.

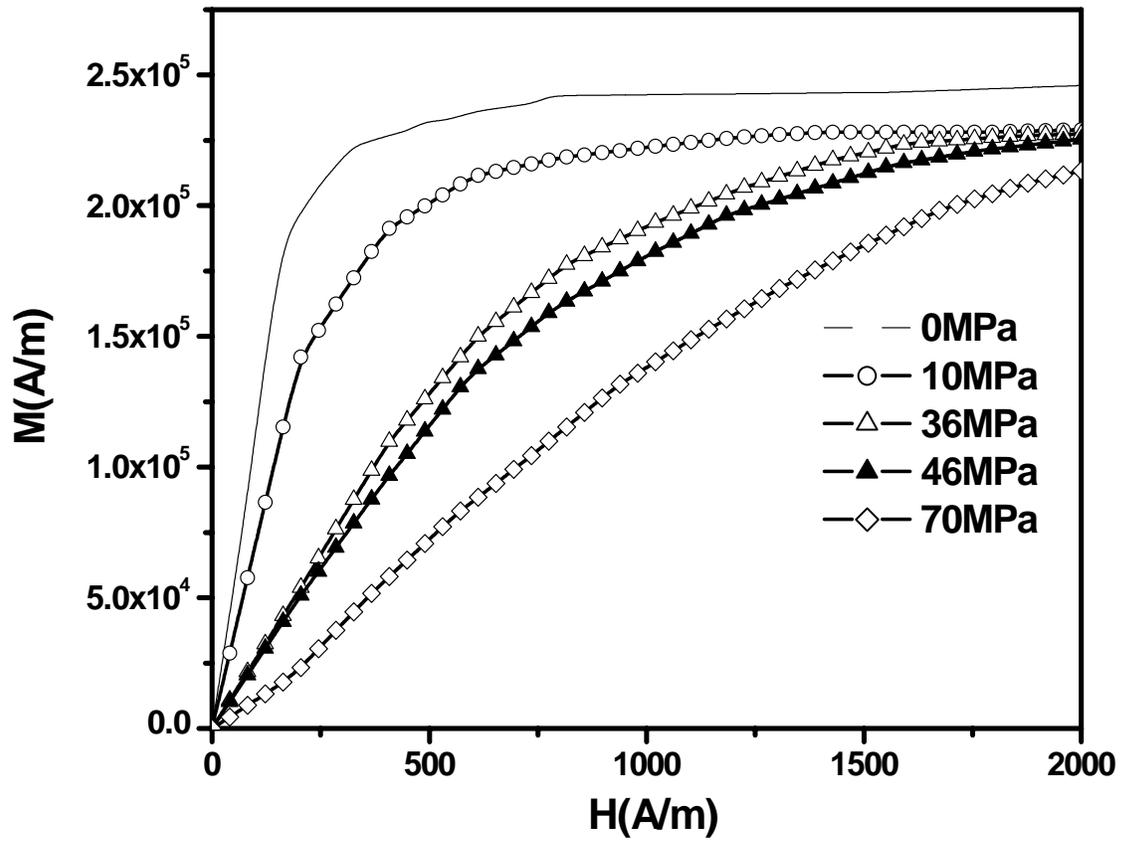

**Fig. 8. B. Kaviraj et al.**

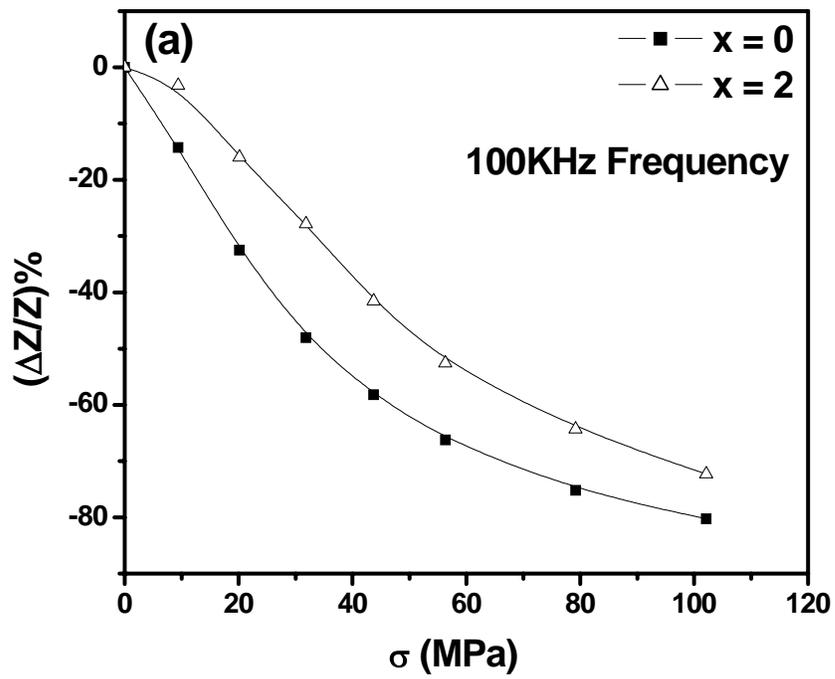

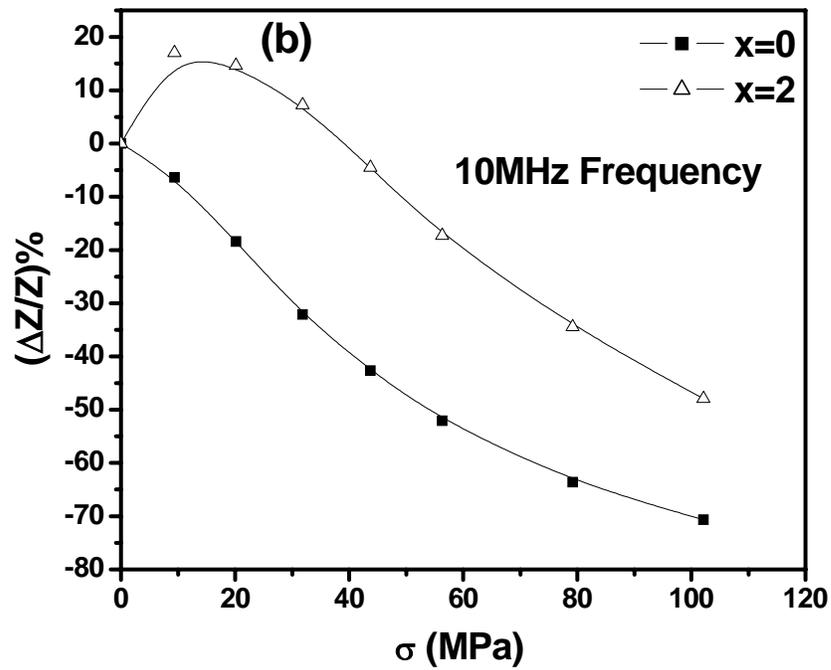

Fig. 9. B. Kaviraj et al.

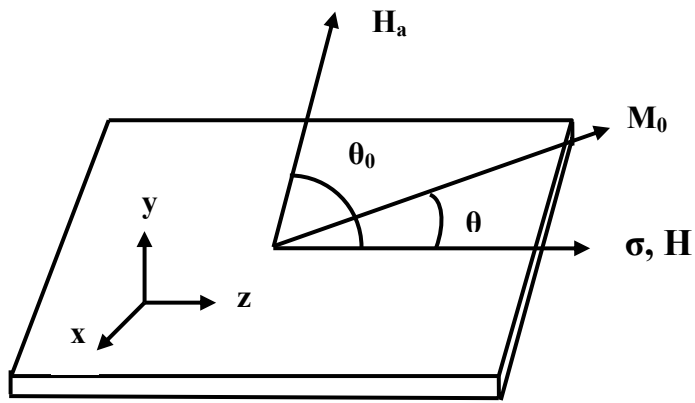

**Fig. 10. B. Kaviraj et al.**

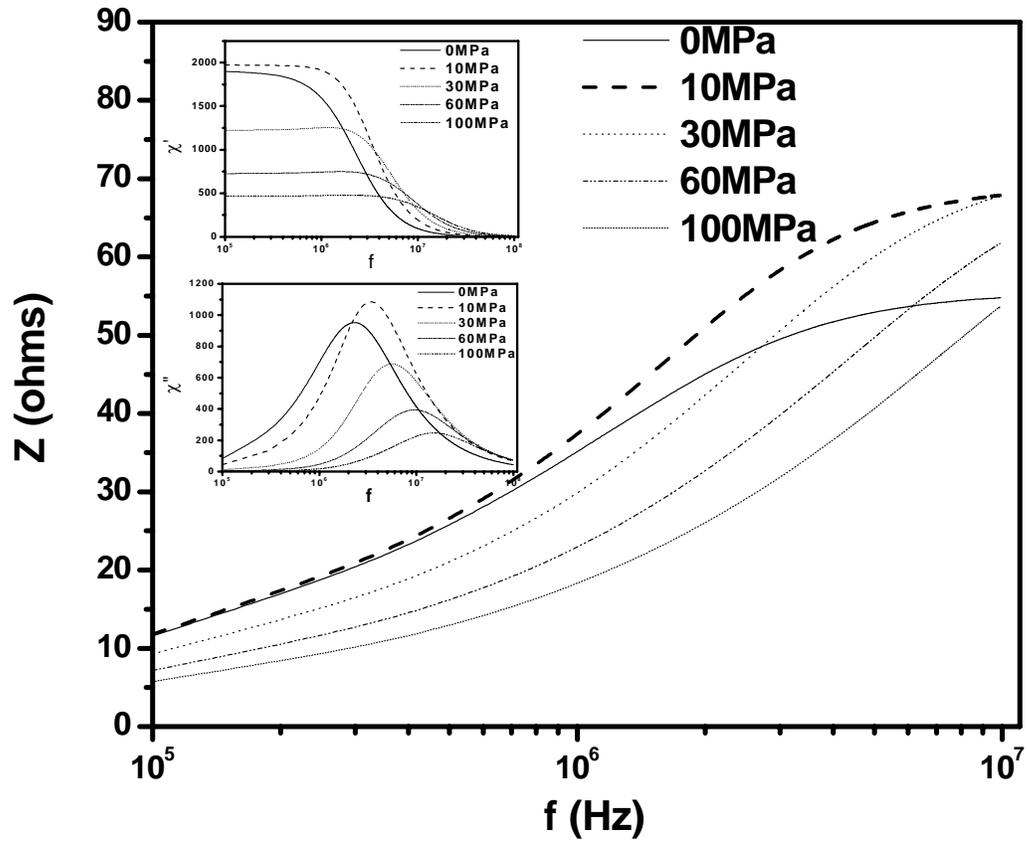

**Fig. 11. B. Kaviraj et al.**

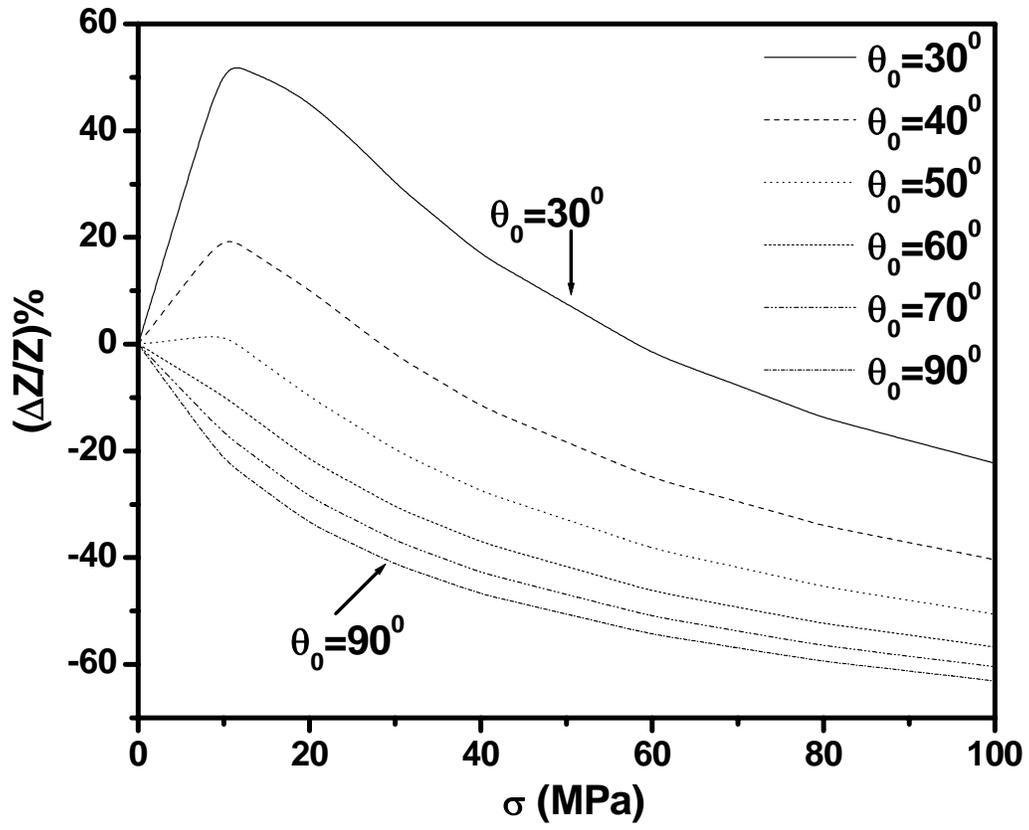

Fig. 12. B. Kaviraj et al.

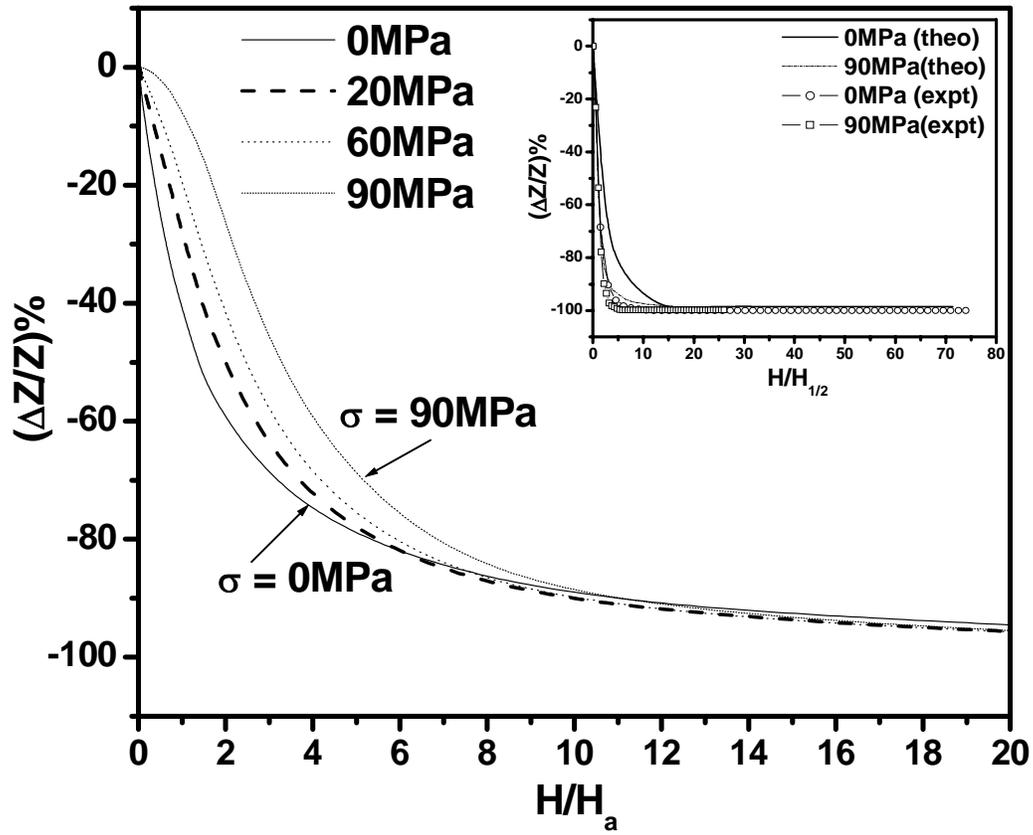

Fig. 13. B. Kaviraj et al.

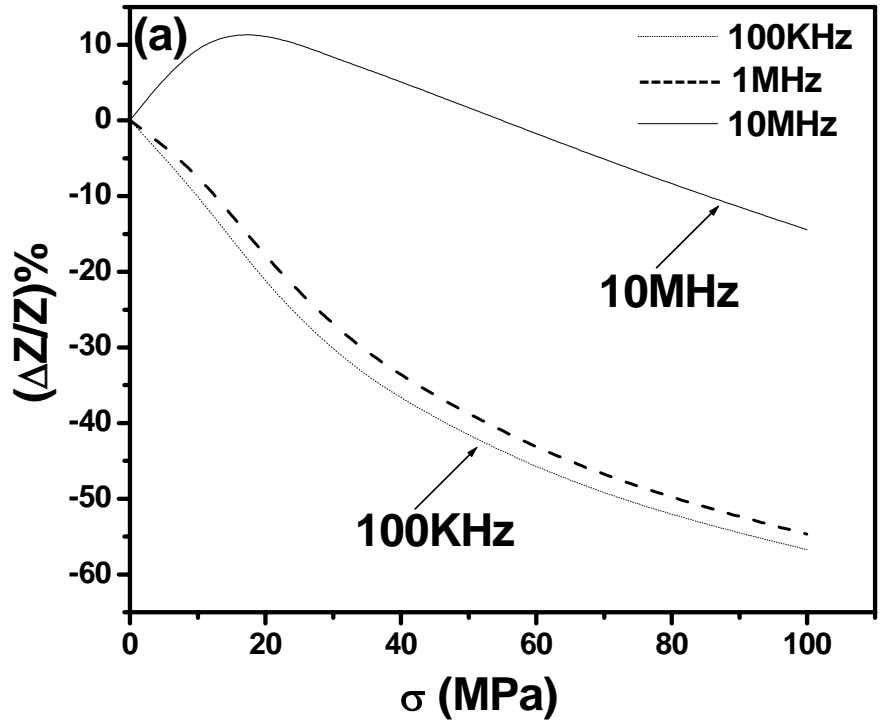

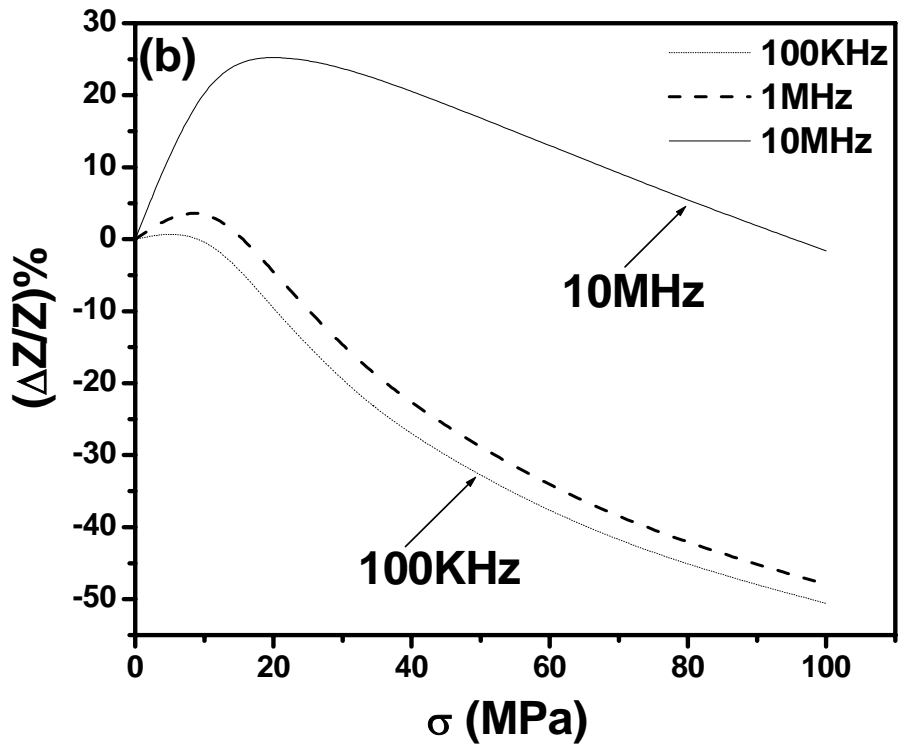

Fig. 14.  B. Kaviraj et al.